# Human Factors Considerations in Satellite Operations Human-Computer Interaction Technologies: A Review of Current Applications and Theory

David G. I. Heinrich, Ian McAndrew and Jeremy Pretty

Department of Human Factors, Capitol Technology University, Laurel, Maryland, USA

## ABSTRACT

*Satellite operations are a subset of remote operations that draw similarities with remotely piloted aircraft (RPA) and uncrewed aerial vehicle (UAV) operations. Increased research into boredom, complacency, habituation, and vigilance as they relate to satellite operations is required due to a lack of prevalence in the literature. Circadian rhythms, crew resource management, and shift work dynamics may exacerbate complacency-driven automation bias and social loafing errors in satellite operations. This overview of theory and applications aims to specifically focus on satellite operations literature within human factors research to identify areas requiring an expansion of knowledge. The human-in-the-loop commonality enables human factors lessons to be passed to satellite operations from unrelated sectors to mitigate catastrophic human error potentially. As such, this literature review details the need for increased research in satellite operations human factors.*

## 1. INTRODUCTION

The United States (U.S.) Department of Defense (DoD) conducts command-and-control (C2) of space assets with a human-in-the-loop (HITL) 24-hours a day, 7-days a week, whereby multiple teams operate in shift work patterns that go beyond regular day-shift hours. Acquisition program offices often place initial developmental and funding emphasis on the satellite system-of-systems space segment, also known as the spacecraft, due to complexity and lack of maintenance capabilities once in orbit. Complexity engineered into spacecraft systems leads to the development of autonomous machines that only need occasional human intervention, thus decreasing overall stress on the operator [1]. Operators must be vigilant to deter the risk of near-peer adversaries and on-orbit anomalies that can degrade or permanently end mission capability. Spacecraft and ground architecture autonomy create the potential of increased operator complacency risk, which may be further exacerbated by circadian rhythm deficiencies due to shift patterns in windowless secure operations centers [2, 3].This study aims to identify the current state of the literature in human factors as it pertains to satellite and remote operations. This paper details the background, significance, current applications, and theories pertaining to complacency, crew resource management (CRM), and human dynamics in remote operations environments.  This review of current applications and theory explores how the literature has not fully fused aviation lessons learned with uncrewed spacecraft operations to combat Gordon Dupont's human factors "Dirty Dozen"[4].

DOI: 10.5121/ijmit.2021.13303           23

## 2. BACKGROUND

According to the U.S. Department of Health and Human Services, shift work is a staffing method that ensures the entirety of a 24-hour day has a constant worker presence for the period beyond regular daylight hours [5]. Shift work can range from three 8-hour shifts to extended shift periods consisting of rotating 12 to 24-hour durations seen in the law enforcement, medical, military, and transportation sectors [5, 6]. Shift work patterns incorporate rotating work-rest cycles when multiple team members are used to fulfill long-term work schedules. Work centers benefit from a constant presence of workers, and the workers themselves also tend to benefit from working shift work patterns outside of traditional day shift hours. Workers may favor shift work out of necessity due to increased pay, work-life balance, educational advancement opportunities, and less perceived supervision on later shifts [5].

Shift workers in satellite operations centers rely on complex systems which create autonomy in space and ground system architectures. Autonomy consists of built-in pre-coordinated reactive measures that ensure built-in internal redundancy keeps the system operating during a malfunction, with no loss of service to the user[1]. Depending on the manufacturer, multiple subsystem units provide redundant failover options, either automatically or manually, with operator intervention. Due to the lack of maintenance actions available to the satellite in orbit, engineered autonomy ensures successful failover to the redundant unit. Barring extraordinary circumstances, most spacecraft can continue providing service by performing autonomous failover measures, while manual failover actions could result in temporary loss of service due to the potential for a delayed human response [7]. Once the spacecraft has triggered an alarm indicating an active non-nominal event or anomaly, satellite operations center (SOC) operators are expected toverify that the proper autonomous failover successfully executed immediately. After safety of the spacecraft is confirmed, operators must then research and diagnose the issues leading up to the failover and implement procedures to return the satellite to nominal operations[7].

Spacecraft or satellite autonomy ensures that the satellite operates with minimal operator intervention[8]. Spacecraft location data called ephemeris is transmitted to the ground antenna, which then is transferred to a data processing site where the data is analyzed to ensure station-keeping accuracy within the spacecraft's proper orbital location. Once ephemeris is validated, an updated maneuver plan is commanded to the satellite, which then must be ingested and computed by the satellite onboard electronics. Once the onboard electronics validate the burn plan, the satellite can continue to operate until the next scheduled ephemeris cycle is due. Satellite operators often work entire shifts only conducting health and safety contacts with the spacecraft to include ephemeris uploads and verifications. Above routine station-keeping requirements, operations personnel maintain a monotonous and stagnant posture. The tedious nature of SOC operations may create the potential for complacency-induced errors. These errors range in severity from inaccurate documentation of shift actions to catastrophic complacency-induced errors resulting from procedural mistakes that can take the satellite out of mission or render the satellite entirely inoperable[9]. Lessons learned from complementary aerospace sectors and unrelated fields need to be evaluated to understand the potential benefits of transferring mitigation techniques into satellite operations methodologies.

Human factors considerations in the uncrewed aircraft systems (UAS) and remotely piloted aircraft (RPA) sectors complement the human factors considerations of satellite operations due to the level of shift work required, level of autonomy involved in their respective systems, and the remote nature of the operations environment. The U.S. DoD employs over 11,000 UAS systems worldwide [10]. Conversely, the U.S. Space Force (USSF) operates over 200 satellites [11], while the USSF 18th Space Control Squadron monitors over 3,200 active satellites [12, 13].

While there is a vast difference between the amount of active UAS assets versus satellites employed by the DoD, both sectors may benefit from a transference of human factors knowledge and lessons learned. Furthermore, the medical, maritime, and transportation sectors can all be leveraged for complementary human factors methodologies.

## 3. SIGNIFICANCE OF STUDY

U.S. DoD single unit satellite acquisitions can cost taxpayers upwards of over $3 billion, with ground segment lifecycle costs for development, operations, and sustainment reaching $6 billion [14]. The USSF Global Positioning System (GPS) III Follow-On upgrade finalized in 2018 will end up costing taxpayers $7.2 billion for a total of 22 satellites [15]. However, the return on investment aims to continue to contribute to the $1.4 trillion in economic benefits since 1988 [16]. Due to satellite programs' proprietary and often confidential nature, little may be known of the full impact of human error in billion-dollar procurements. Nonetheless, human error has been exhibited in space operations [17]. Incidents of invalid software loads, incorrect coordinates set before launch, and other human errors have led to total mission loss [9, 17]. Transference of lessons learned between sectors may establish common knowledge, which then may leverage best practices between sectors, thus potentially reducing the overall risk of human factors-related mishaps.

Knowledge transfer between aerospace and non-aerospace sectors will ensure that human factors lessons learned are analyzed and incorporated into the development of costly governmental and commercial investments. Human error must be further engineered out of remotely operated space systems to preserve capabilities, such as GPS, that have become vital to everyday life[16]. Steps taken during SOC development to mitigate complacency-induced errors will build a foundation of error prevention that will have lasting impacts on the future of satellite operations. This review of the literature expands upon the works of Dupont, Hawkins, and Reason to ensure the field of satellite operations is considered in the human factors conversation[4, 18, 19]. Without careful consideration, the proliferation of space-based technologies may suffer a repeat of comparable historic human factors related aviation mishaps as seen in the 1977 Tenerife Airport Disaster [18], which would only serve to stifle progress if not for human factors lessons learned.

## 4. NATURE OF STUDY

This research paper comprises a meta-analysis of the current literature on automation, autonomy, complacency, shift work, and shift patterns in the remote operations aerospace sector. SOC operations are not unique regarding human factors as there are other human-centric sectors where lessons learned may be considered: healthcare, maritime, and the transportation sectors. Satellite operations deal with similar human factors issues as seen in the UAV sector. Thus, a meta-analysis of qualitative research was selected due to the potential classified nature of military satellite operations. This study remains in the unclassified domain through analysis of existing open-source data.

## 5. CURRENT THEORY AND APPLICATIONS

### 5.1. Theoretical Foundations

This research paper comprises a literature review of foundational research,which will be expanded upon via future studies and analyses. The theoretical framework of this research is based on the existing body of knowledge of aerospace and non-aerospace concepts and theory. The theoretical foundation of this research paper revolves around the concept of Gordon

Dupont's "Dirty Dozen"[4], Frank Hawkins software, hardware, environment, liveware, and outside liveware (SHELL) model [18], James Reason's "Swiss cheese" Model of Accident Causation [19]. These foundational concepts will be verified in future studies via qualitative observational research.

### 5.2. Human Factors Concepts

#### 5.2.1. Complacency

According to Årstad and Aven, complacency is "unintentional unawareness [which can only be diagnosed] in hindsight, from a distanced perspective" [20, p. 115]. The National Aeronautics and Space Administration (NASA) defines complacency as "overconfidence from repeated experience on a specific activity, complacency has been implicated as a contributing factor in numerous aviation accidents and incidents" [21, p. para.1]. Parasuraman et al. [22] referenced the prevalence of complacency throughout multiple aviation accident investigations lending credence to the importance of Dupont's "Dirty Dozen" [4]. Prinzel cited crew complacency as often being "a contributing factor in aviation accidents" [23, p. 4]. Merritt et al. referenced "complacency, or sub-optimal monitoring of automation performance, [as being] cited as a contributing factor in numerous major transportation and medical incidents" [24, p. 1]. Overconfidence in highly reliable automated systems often leads to complacency issues [22].Complacency is often cited alongside boredom and procrastination[21, 23] . Prinzel found that those with low self-efficacy "suffered automation-induced complacency" and operated significantly better when working in high workload environments[25, p. 13]. Furthermore, Prinzel [23] also found that pilot workload over-saturation can increase, leading to an overburdened cognitive load. Conversely, boredom has been shown to increase when the operator defers to the machine due to the repetitive nature of automated tasks in a cognitively low-demand environment [26].

Aviation human factors incidents involving pilot complacency have been attributed to a failure to adequately correct automation errors [22, 24, 27]. Merritt et al. [24] highlighted that complacency could manifest due to a person's inability to comprehend the occurrence of an error or exhibit a prolonged response to an error or stimuli. Beyond prolonged reaction to an automation failure, failing to act may be attributed to both commission and omission errors [23, 24, 28].Errors of commission happen when the HITL makes a mistake or error due to incorrect decision-making. Errors of omission occur when the human does nothing when they should have[24, 29].

Complacency research pertaining to maritime shipping operations identifies similar outcomes, which serves to highlight Dupont's "Dirty Dozen" [4, 30]. Attempting to fill gaps in maritime research, Bielic et al.[31]studied technology, leadership, management, and self-induced complacency during their study of complacency in maritime accidents. Bielic et al.[31]cite research by Turan et al.[30], who found that over 80% of maritime shipping incidents were attributable to human or organizational error, of which 6% could be attributed to complacency. The research referenced an overreliance on automation as a leading cause of technology complacency. Furthermore, leadership complacency may increase the risk of complacency in the same manner as poor team dynamics, toxic hierarchy, and steep authority gradient, as seen in aviation sources [31]. Leadership complacency occurs when the leader or manager possesses an inadequate leadership style. Workers may become apprehensive about going totheir leadership out of fear of being ignored or potential mistreatment[31]. Alternatively, leadership complacency may occur when leaders are not professionally challenged during critically sensitive moments. During the events leading up to the Chernobyl disaster, workers were discouraged from having a questioning attitude, and the overall lack of open communication resulted in increased loss of life due to the amount of time it took the workers to react to the situation[31, 32].

Satellite operations assets rely on a system-of-systems to provide autonomy due to the inherent complexity of satellite electronics and the inability to conduct on-orbit servicing. Due to this complexity, human satellite operators and the spacecraft, or autonomous agent, function as a team [33]. While not explicitly stated, the concept of human operators and autonomous agents working together points to the idea of crew resource management (CRM), where both entities operate as a crew working toward a common goal [34, 35, 36, 37]. Communication between the autonomous agent and humans exists within the environment, as seen in the human factors SHELL model where the liveware, hardware, and software work together [18].Complacency may occur due to automation bias when humans defer decision-making and authority to more complex autonomous agents [38]. Lyell and Coiera [38]found that task complexity and difficulty play a significant role in automation bias due to the level of difficulty present when monitoring and interpreting automated aids in the healthcare field.

### 5.2.2. Boredom

Boredom is a precarious state of awareness that has been shown to lead to errors of omission in automated systems [39]. Pope and Bogart coined the term "hazardous states of awareness" to encompass the concepts of boredom and inattention [39]. Prinzel et al. [23] highlighted the need for more boredom-related research. The Aviation Safety Reporting System(ASRS) provides a multitude of evidence that complacency and boredom are linked, leading researchers to conclude that complacency is tied directly to boredom[23, 39].The work of Prinzel et al. [23] pointed toward individuals with high-complacency-potential to suffer more from boredom than individuals with low-complacency-potential. Prinzel et al. [23] admitted that additional research using actual pilots in future studies is needed due to the possible psychological attributes or personality differences from subjects studied in their research. The work of van Hooft and van Hooff bolstered earlier findings, calling for methods to increase attention and add autonomy to help alleviate the risk of error during operationally tedious phases of work to mitigate human factors errors [40].

### 5.2.3. Habituation

Habituation is commonly associated with boredom and complacency due to varying stimuli within a human's daily life or work environment. A classic example of habituation can be taken from Kim and Wogalter's study [41], highlighting the use of safety warnings within the work environment. Static standardized warning signs displayed around industrial work centers were studied, and researchers found warning signs were ineffective at keeping attention in the long term [41]. Kim and Wogalter [41] concluded that signage needed to be changed, enhanced, or embellished to increase stimuli over the long term to increase human attentiveness to safety warnings. Decrements in sensation in response to a given stimulus may occur over time due to a feeling of increased control. The human begins to anticipate outcomes, and their subsequent risk tolerance grows while their awareness decreases, leading to increased complacency concerning the identified risk [42]. Sensitization via training or providing an increased stimulus to elicit a reaction are ways to combat habituation [43]. A constant or continuous stimulus can become adaptable for the human. Constant noise can become easy for the human brain to filter out due to the volume and consistency of the noise. The brain will only react when the noise changes to a different tone, pitch, or volume [44].

### 5.2.4. Condition Monitoring

Condition monitoring aims to increase knowledge of the status of a system through automated vigilance [45]. Machine condition monitoring dates to 1924 and initially dealt with motor faults. Condition monitoring then progressed to the use of vibration in the 1950s. Modern advancements have led to permanent radio frequency sensors to relay data from multiple points of interest to understand system health and to increase the mean time between failures [46]. Dadashi et al. [47] highlighted automated remote sensing issues as a contributing factor in the 1979 Three Mile Island and 1994 Milford Haven, U.K., Texaco Refinery explosion incidents. The Three Mile Island nuclear reactor used hundreds of separate alarms to convey status with key indicators obscured to the operator [48]. Within the beginning stages of the Three Mile Island incident, over 100 alarms were triggered, causing the operators to react to an overload of information. In adequate training and deficiencies in condition monitoring protocols were cited as contributing human factors in the incident [49].

### 5.2.5. Alarms and Warnings

Satellite systems operate on a crewed shift concept where on-shift operators monitor, diagnose, and remedy system issues via telemetry and alarm protocols [50]. Alarms in crewed and uncrewed systems alert the operations crew of a problem with onboard components or an issue in the current operational situation. Theories on interpreting and understanding the status of a crewed spaceflight system have been covered in the literature dating back to the 1950s [51]. Bamford et al. [51] stated that an appropriate response would be warranted for any given stimuli, with the level of operator response being improved upon through training and indoctrination. Lange et al. [52] presented findings consistent with a need for training and indoctrination of crews in any field. They [52] found that users of a system need to understand the monitoring device and its associated functionality to use the device correctly and efficiently.

Alarm priorities assigned during development are imperative to mitigating confusion. According to Laughery and Wogalter [53], three criteria must be considered when prioritizing warnings: likelihood, severity, and practicality. The Texaco Refinery explosion was attributed to several factors, including poor alarm management design and prioritization [54]. During the 11 minutes leading up to the Texaco Refinery explosion, there were 275 active alarms present, with 87% of the active alarms considered a high priority [54, 55]. The Engineering Equipment and Materials Users Association (EEMUA)[56]recommends that only 5% of alarms be regarded as high priority, 15% medium, and the remaining 80% of a system's alarms should be reserved for low priority alarms. Foong et al. [57] analyzed alarm prioritization in oil refineries and agreed with the EEMUA [56] by confirming a best practice prioritization methodology of 5% for emergency priority, 15% for high priority, and 80% for low priority alarms.

### 5.2.6. Communication Human Information Processing (C-HIP)

Industrial risk research has led to the concept of Communication Human Information Processing (C-HIP), whereby an alarm flows from the machine to HITL, who must then react appropriately to resolve the alarm [58, 59]. While C-HIP is aimed toward industrial work center warning human factors, the concepts can be expanded to other industries where warning and alarms are employed [60]. C-HIP consists of several serial steps starting at the source of the warning and ending at a desired human behaviour [61]. The source of the alarm can originate from the software, hardware, or liveware [18, 60]. In a remote system, the channel presents the alarm either auditorily, visually, or both. The operator of the remote system is the receiver. The receiver category is the "human information processing section of the C-HIP model", where human factors of attention, comprehension, attitudes, beliefs, and motivation reside[58, p. 314]. The receiver or HITL must then process the alarm and act accordingly to mitigate potential system degradation [58, 59].

## 5.3. Shift Work

In 1997 Gordon Dupont presented his "Dirty Dozen errors in maintenance" consisting of the 12 most common causes of error judgment in aircraft maintenance: "lack of communication, complacency, lack of knowledge, distraction, lack of teamwork, fatigue, lack of resources, pressure, lack of assertiveness, stress, lack of awareness, and norms"[4, p. 42]. Dupont's common causes have been used extensively in the aviation maintenance human factors arena, yet principles identified by Dupont extend to other sectors. As with aviation maintenance, multiple industries such as medical, transportation, services industries, and aerospace operate using shift work in varying degrees [6, 62, 63]. Sectors that employ shift work tactics operate with the same risk of human factors consequences highlighted by Dupont, yet his human factors principles have not fully extended outside of aviation literature.

Sleep patterns among shift workers in the crewed space arena experience human factors issues highlighted in Dupont's "Dirty Dozen"[4]. Mizuno et al. [62] studied Japanese International Space Station flight controllers' sleep patterns. The Japan Aerospace Exploration Agency used a 3-shift-per-day schedule "from 8 a.m. to 5 p.m. (day shift), from 4 p.m. to 1 a.m. (evening shift) and from midnight to 9 a.m. (night shift)"[62, p. 3]. Their research found that night shift workers encounter the most challenging fatigue and sleep issues. Night shift workers were shown to have a higher prevalence of shift work sleep disorder (SWSD) than day and evening shift workers, where insomnia symptoms and decreased working competency were observed [62]. Åkerstedt [64] also found decreased working competency and insomnia symptoms in those that suffer from SWSD. SWSD has been shown to increase the prevalence of the same issues highlighted by Dupont [4] with a higher prevalence of stress and fatigue as a result of atypical shift patterns, specifically as a result of working the night shift [2].

SWSD has been correlated with circadian stress in night shift workers [62]. Sleepiness among night shift workers may lead to wakeful fatigue, which has been shown to be more prevalent in night shift workers than day shift workers [64]. Research in the public health sector has correlated circadian rhythm disruption with SWSD, which has hindered human performance during the night phases of a work pattern [28]. Sleep-wake cycles may become disrupted when initially entering a night shift pattern in the same manner as seen with jet lag. Pharmaceuticals and therapeutics have been studied to combat the effects of jet lag with varying levels of success. Jet lag, also known as circadian desynchrony, has been shown to produce the same symptoms as SWSD with "disturbed sleep, daytime fatigue, decreased ability to perform mental and physical tasks, reduced alertness, and headaches" caused by circadian stress [18, 65, p. 221].

SWSD and circadian rhythm issues have been shown to lead to mood issues among those who work night shifts due to excessive sleepiness, fatigue, and trouble sleeping [66]. Walker et al. [66] detailed the correlation between circadian rhythm disruption and poor mental health due to shift work. Jehan et al. [2] also found an increased prevalence in sleep disruption among medical workers on the night shift versus those that work during the day. As a result of sleep disruption, poor mental health and anxiety may become exacerbated by shift work among health care workers [67, 68]. Night shift work has been associated with depression due to circadian misalignment [69]. All instances of poor mood or mental health have been partially attributed to either sleep quality, sleep quantity, sedentarism, and poor diet choices during night shift work [2, 66, 68, 69].

Shift work research points toward late-night shifts, between 1 a.m. and 8 a.m., as a prime time for medical and performance-based human error occurrences [70]. Circadian rhythm and SWSD tend to make night shift work more precarious [2, 62, 66, 71]. Inadequate sleep before working a night shift, even 1-2 hours less, can drastically increase the potential for human error [70]. Mizuno et al. [62] suggest that some workers are better suited for night shift work than others. Operations could benefit from using individuals better adjusted to night shift work to mitigate human error. Attention to identification methods and policy formulation have been highlighted to help reduce human error during vulnerable shift time frames [70]. Hege et al. [63] referenced a need for a comprehensive review of commercial delivery drivers' regulations and operational conditions to enact regulations to prevent human error in the transportation industry.

Multiple human factors incidents have been attributed to night shift work due to a "30% increase in human error incidents on night shift relative to morning shift" [28, p. S88]. The most notable shift work-related incidents are the Chernobyl nuclear reactor meltdown, 1985 Davis-Besse nuclear reactor incident, Exxon Valdez oil spill, Space Shuttle Challenger accident, Three-Mile Island nuclear reactor incident, and the1985 Rancho Seco nuclear reactor incident [28]. The Chernobyl nuclear power plant incident, cited as the worst nuclear disaster ever, occurred at 1:23 a.m. The Davis-Besse reactor incident started at 1:35 a.m. due to night shift operators not following correct protocols. The Exxon Valdez oil spill occurred just after midnight when the vessel collided with Bligh Reef. "The failure of the third mate to properly maneuver the vessel, possibly due to fatigue and excessive workload" was a contributing factor cited in the National Transportation Safety Board investigation [72, p. 2]. The Space Shuttle Challenger incident investigation highlighted shift work, and overwork issues at the Kennedy Space Center as contributing factors in the accident [73]. Three Mile Island took place between 4:00 a.m. and 6:00 a.m. when night shift workers failed to recognize they had a problem. The 1985 Rancho Seco incident occurred at 4:14 a.m. when errors of commission and omission resulted in a delay to regain control of the site[28]. Mitler et al. [70] state that human factors leading up to the referenced incidents may not have been the root cause, but human factors played a significant role in how the incidents occurred. In the Exxon Valdez Oil Spill, Three Mile Island, Chernobyl, and Space Shuttle Challenger disaster: operator fatigue was cited as a critical reason for the mishap due to lowered alertness, inattention, and delayed reaction [70, 72].

### 5.4. Crew Resource Management

Crew resource management (CRM) was first implemented into aircrew training in 1979 after a series of high visibility aircraft accidents [36]. CRM is a concept that consists of training aimed at aircrews to increase communication and teamwork of those involved in all phases of flight. Teams outside aviation have since incorporated CRM training and concepts to decrease the chance for human error. While CRM implementation has been ongoing, human error continues to be prevalent due to a lack of support of recurring CRM training [74]. Salas et al. [74] argue that straightforward methods have not been translated from science to the education system, and only through a standardized methodology will CRM become more effective. CRM has been defined as using all resources, information, and people in the loop to conduct safe operations [75, p. 4]. The 1994 Operation Provide Comfort fratricide incident between a patrolling friendly U.S. Air Force fighter aircraft and a U.S. Army Black Hawk transport was notably caused due to poor CRM of the Black Hawk crews, F-15 pilots, Airborne Warning and Control Systems crews, and the general lack of communication of proper rules of engagement before mission execution on the part of all parties involved [75, 76]

Aviation is not the only sector that references CRM implementation: maritime, medical, and remote operations have been referenced in needing effective CRM. Medical operating room environments contain hierarchy gradients and adverse workplace cultures that often create ineffective teamwork barriers. Hierarchy gradients have been cited as causing human error in medical situations [77].The medical sector has started to pursue more open communication through the exact CRM implementation that the aviation industry uses to prevent human error [77]. Pilot studies in the medical field have pointed to the usefulness of simulation training events to highlight and reduce CRM hierarchy-related communication errors [78].

In an extensive study on medical hierarchy gradient, 363 medical employees were surveyed about safety reporting. Siewert et al. [79] highlighted that senior employees were more likely to speak up than their lower-ranking counterparts due to the following human factors:

High reporting threshold (i.e. [sic], uncertainty about one's observation), reluctance to challenge someone in authority, lack of being listened to, fear of retribution, fear of disrespect being expressed, individuals creating an uncomfortable work environment (toxic captain), shy personality, responsibility within a team, and lack of language skills. [79, p. 796]

Authority or hierarchy gradient was the second most reported issue after the high reporting threshold [79]. Physician trainees have been identified as the most vulnerable to hierarchy gradient in the medical profession, yet they may be the most up to date with current information. In 2018, Siewert et al. [79] stated medical information would double every three years, and by 2020 medical information would double every 73 days. The study further highlighted the need for increased CRM training, citing positive human factors changes in the aviation industry due to the effective employment of CRM [79].

CRM training levies an emphasis on communication [34, 35, 37, 74]. CRM indoctrination focuses on members working as a team rather than a group of separate individuals working together [34]. CRM training operates with the acceptance that there will be a human error, and the proper way to react and mitigate the error is through effective CRM [80]. Gross et al. [37] conducted a meta-analysis of 1,037 medical publications on CRM; they found communication, situational awareness, leadership, teamwork, and decision-making as the top five CRM topics in healthcare literature. However, only a fraction of the publications identified provide an "explanation sufficient for replication" when creating CRM training materials [37, p. 6]

The findings of Gross et al. [37]are consistent with that of Salas et al.[74], who highlighted that CRM training has not been employed using consistently actionable or repeatable training methods that can be recreated throughout training. Gross et al. [81]further analyzed constructs in CRM training to make a case for efficiency with 15-minute training sessions, versus 2-hours up to full day training sessions. They expanded upon the theme of CRM training inefficiency with 129 participants to understand if periodic micro-training sessions were more efficient in the long-term employment of CRM principles. Gross et al.[81]found newly acquired CRM principles were able to be recalled weeks later, leading to the conclusion that micro-training sessions may be a valid method of training CRM in time-constrained professional communities. The science of training regarding how information is "designed, delivered, and implemented can greatly influence its effectiveness" [74, 81, p. 3].

Professionals in the medical field see real-life emergencies where pressure is either perceived or realized every day. Time-sensitive and highly stressful medical events have led to novel methods in CRM training and employment. The "10-seconds-for-10-minutes principle" was created in the medical sector to increase situational awareness in high-stress scenarios [82]. Rall et al. [82] cited a speed and accuracy trade-off where perceived time pressure can lead to error; thus, slowing down can help mitigate error by stopping all work, analyzing the problem, gathering facts, planning, and distributing the team's workload in manageable quantities. "Teams who took a fraction longer at the start of an emergency (to assess the situation and brief the team) performed better" [83, p. 95]. Gross et al. [81] developed an alternative to the 10-seconds-for-10-minutes principle called the Team Check protocol, where the team asks similar questions: what, how, and who. Both methods slow the team down to increase situational awareness in high-stress scenarios [81, 82].

## 5.5. Remotely Piloted Aircraft

UAV and crewed aviation suffer the same human factors issues, yet mishaps occur one to two orders of magnitude greater than during crewed flight [84, 85]. According to USAF, "Class A mishaps are currently reported any time an incident results in (1) $2 million or more in damage to the aircraft, (2) a fatality or permanent disability, and/or (3) destruction of the aircraft" [86, p. 2]. Furthermore, the USAF considers an aircraft destroyed when the aircraft is rendered unrepairable and cannot be returned to service [86]. Remotely piloted aircraft (RPA) statistically suffered more Class A mishaps and were destroyed more than piloted aircraft during every period recorded [86]. Inconsistent training, non-standardized pre-departure briefings, and poor CRM interactions with air traffic control have all been identified in the ASRS as causal factors in UAV mishaps [87].

Crewed aviation incidents involving fatigue have been cited in 7%, 4%, and 7.8% of accidents in civilian aviation, U.S. Army aviation, and USAF aviation, respectively [88]. Alternatively, 13.85% of aviation accidents were attributed specifically to RPA mishaps [89]. Gaines et al. highlight fatigue as a human factors incident amplifier [89]. While the RPA HITL is safely on the ground during a mission, long-endurance high-tempo RPA flights have led to shift work implementation, which has been shown to carry fatigue-induced consequences [2, 64, 90]. Military RPA operators are prone to fatigue due to the slow-paced monotonous nature of RPA operations [91]. Pedersen et al. [92] called for increased automation to regulate workload to mitigate fatigue in remote operations. Furthermore, USAF RPA operations centers have addressed ergonomics considerations directly related to operator fatigue by addressing: "Climate control, restroom breaks, ergonomic design, and equipment" issues [93, p. 62]. Addressing living condition issues may benefit both the operator and the mission by ensuring preventable human factors issues are mitigated before they manifest through thoughtful operations center design [93].

Tvaryanas and MacPherson found a "do more with less" [94, p. 460] attitude resulted in less-than-optimal personnel levels when researching RPA crew shift patterns. Insufficient personnel levels increase the need to work longer hours, resulting in reduced recovery between shifts worked and an increase in the potential of fatigue-related issues [94]. Fatigue can be exacerbated by a sustained high operations tempo, increasing symptoms associated with SWSD [95]. Multiple studies conducted by AFRL recommend crew rest, regular exercise, proper nutrition, blue light standardization [96], and the application of "science-based shift scheduling techniques when developing manpower requirements and developing duty time and crew rest requirements" [95, p. 33]. Tvaryanas and MacPherson recommend mitigating fatigue in 24/7 operations using multiple crews in multiple time zones where operations would always be conducted during the local day shift [94].

Inadequate or insufficient training has led to RPA human factors mishaps [89]. RPA operations consist of automated human-system integration (HSI) systems that drive increased training requirements [97]. Increased automation has been shown to decrease sensory cueing, resulting in detrimental decreases in situational awareness during RPA operations [98, 99]. RPA operations may benefit from CRM training to mitigate HITL fatigue and vigilance-induced error [34]. RPA operations that do not adequately employ proper CRM create issues with air traffic control and crewed aircraft [35]. Loss of communications has been an issue when incorporating RPAs into the National Airspace System (NAS) [87]. According to Neff, "in order to ensure the safety of all operators in the NAS, CRM should be part of an overall safety management system implemented and practiced by all operators and overseen by the [Federal Aviation Administration] FAA" [35, p. 7]. Salas et al. [36, 74] highlight requirements for practical CRM training while Neff [35] complemented the studies conducted by Salas et al. [36] when he analyzed four RPA CRM-related human factors incidents which could have been mitigated by practical CRM training and implementation.

While human error has been cited in 79% of USAF RPA mishaps [100], loss of human life due to RPA mishaps is negligible. Operators located in RPA ground control stations contend with human factors issues regarding ergonomically inefficient human control interfaces, which can be error-provoking, complex, and tedious to operate [99]. AFRL found that 92% of Predator RQ-1 operators report "moderate to total boredom" [95, p. 26][101]. RPA operators lack sensory cues afforded to conventional aircraft due to removing the HITL from the physical airframe [99]. On June 25, 2020, a MQ-9A Reaper RPA crashed within one minute of takeoff due to pilot misidentification of the flap and condition levers. Instead of reducing flap position after takeoff, the pilot reduced the condition lever, cutting off fuel flow to the engine [102]. The MQ-9A mishap highlights the need for increased human factors development in RPA HSI systems. Prichard [102] highlighted incorrect equipment operation, fixation on the wrong aspects of the situation, inadequate location, and lever color as design and training deficiencies, increasing the potential of future mishaps.

## 6. CURRENT APPLICATIONS

### 6.1. Satellite Operations

Contrary to aviation, the subject of satellite operations human factors is limited in the literature. The USSF produces most literature about satellite operations due to the number of satellites operated by the military branch. Operating multiple satellite constellations safely and reliably is only possible due to organizations functioning in a highly reliable manner [103]. According to Schubert et al. [103], highly reliable organizations start with effective human management and trust. Reason's "Swiss cheese" model [19] depicts management and culture as contributing factors in an accident. Beyond management, the HITL must work with various control interfaces, interpret multiple information sources, and respond correctly to mitigate error [104]. Satellite anomalies occur due to hardware and software malfunctions on-orbit stemming from space weather, orbital debris, hostile actions, electromagnetic interference, and operator error. Operator errors made during the command and control of spacecraft may create an anomaly if the system does not have the correct failsafe mechanisms in place [7].

Space assets have a long history of human error. Russian Soyuz, Phobos 1 Mars probe, Russian Mir space station, Mars Pathfinder, the U.S. Space Shuttle program have all had some level of error attributed to the HITL [9]. The best practice to effectively reduce human error is often during the development phase, where "robust and fault-tolerant spacecraft hardware and software" are designed [105, p. 4]. Human error may occur due to time constraints, level of task understanding, inadequate mission preparation or training, and errors of commission or omission [105]. The development of spacecraft systems is not a standardized process due to the difference in proprietary specifications instituted by different spacecraft manufacturers. Due to the unique nature of satellite acquisitions and development, it may be challenging to standardize human error mitigation techniques [9].

## 6.2. Human-System Integration

HSI is the concept of combining the entire system with the HITL. Human-computer interaction (HCI) is a subset of HSI where the HITL uses a computer to interact with the system. The computer is the portion of the system which links the human to the system [90, 106]. Inadequate operator control stations lacking HSI have been cited as a critical factor in RPA mishaps [107]. The UAV sector has been plagued with human factors HCI issues. High levels of cognitive load can occur when attempting to process multiple screens at the same time. The HITL may have to contend with having to enter over 20 inputs to engage autopilot, inadequate feedback mechanisms, and multiple input mechanisms in a cluttered work area to operate an UAV successfully [108]. HSI issues factor in 89% of MQ-1 Predator accidents, while the HSI has been cited as a causal factor in 44% of MQ-1 Predator mishaps [108]. Intelligent software used during HCI development may aid the operator in error mitigation, fault isolation, and operator input prediction [9]. HCI design must be fully developed to ensure the HITL can take over anytime, especially during an emergency, no matter the level of automation or intelligence built into the system: the HITL must always be in control [109].

## 6.3. Automation/Autonomy

Automation carries out tasks generally performed by a HITL in a controlled manner through process implementation [110]. Autonomy creates independence from the HITL; thus, the system decides and implements the next course of action [84]. Automation and autonomy are implemented to remove the HITL from the system so long as no deviations occur [84]. According to the U.S. DoD [84], the core difference between automation and autonomy is the governance of rules: broad for autonomous and prescriptive for automated. Depending on the development and implementation of automation and autonomy in the system, HITL involvement levels may vary according to manufacture specifications.

HITL complacency when working with automated or autonomous systems is possible due to a phenomenon called "social loafing" [111] or "The Ringelmann Effect" [112]. In group settings, "social loafing" may occur when multiple people work together and may defer responsibility to others in the group to accomplish a task versus putting in more effort if they had acted alone [113]. Groups of people working together have been shown to make twice as many vigilance errors versus working alone while adding an additional HITL does not necessarily increase vigilance [114]. Automation bias or the deferring of responsibility to the machine has striking similarities to "social loafing" [114, 115] in that there may be an over-reliance on automation and potential for misplaced trust in the accuracy of the system [110].

Automation complacency has been cited in multiple human factors incidents where automation failed, yet the HITL overlooked critical indicators due to their over-reliance on the system's accuracy [115]. The 1995 Royal Majesty cruise ship accident occurred due to several reasons: failure to follow established procedures, lack of situational awareness, incorrect routing of GPS cabling, and automation complacency [116]. The accident report found that the accident occurred due to the crew's overreliance on the system and a maintenance error resulting in a disconnected GPS cable [115, 116].

## 6.4. Big Data Handling

The FAA handles 16,405,000 flights each year with 45,000 flights occurring each day [117]. Meanwhile, only 11,000 UAS systems [10] and over 3,200 active satellites are employed in the civilian and military sectors [12, 13]. The amount of data accumulated in aviation is immense compared to that of UAS or satellite operations, with "the average flight data collected during a current flight operation [getting] up to 1000 gigabytes" [118, p. 214]. The large data originates from aircraft onboard electronics like those used in satellites [9, 118].According to Oh [118], large aviation data is considered Big Data. Madden states that Big Data is either "too big, too fast or too hard" [119, p. 4] for most existing software tools to process [120]. Larger Big Data quantities may get up to "petabyte-scale collections of data" or 1 million gigabytes, which could accumulate faster than processed, making analysis and decision-making difficult [119, p. 4]. Maritime logistics has seen an increase in the volume of data used in shipping [121].According to Yuen, Xu, and Lam, maritime research of Artificial Intelligence (AI) and Big Data has been directly attributed to cost savings for ship operators due to the effective employment of mined data [121]. AI has proven useful in analyzing Big Data, yet software can pose human factors issues due to visual noise, perception, monitoring issues due to image challenges, and the high-performance requirements necessary for the software to operate, which all may serve to slow down the HITL [118, 122].

Big Data analytics are employed by business managers to better understand cost and efficiency metrics [123]. Big Data has been analyzed by "schema-less databases capable of handling large amounts of structured and unstructured data like documents, e-mail and multimedia efficiently with flexible data models" [123, p. 1]. There are three types of analytics used in Big Data processing: descriptive analytics, where the data shows what has happened; predictive analytics, which can help to forecast what might happen in the future; and prescriptive analytics, which can help figure out how to prompt an event to occur [124]. Large volumes of data in aviation, business, and maritime operations are like that of the telemetry data transmitted by satellites for health and safety status, imagery, and communications [7, 9]. Big Data's usefulness depends on the training of the HITL in the use of intelligent software to analyze the data [125]. Avci, Tekinerdogan, and Athanasiadis [126] reviewed software architectures for Big Data. They found 43 studies covering Big Data and highlighted several applications currently in use [126]. Avci et al. did not find a consensus for the proper or most useful type of software used for Big Data analysis which is consistent with how most proprietary methods are marketed and employed [126]. While Big Data is being explored for use in agriculture, climate change, and remote sensing applications, the literature has a low prevalence of research on how to employ Big Data analytics to combat human error [124].

## 7. INTERPRETATION OF THE LITERATURE

The state of the literature may be incomplete concerning advancements in HSI or HCI development due to a delay in reporting as a result of security considerations. As referenced in the literature review, AFRL publishes most human factors research on RPA and satellite operations. However, the data may be considered outdated or potentially blocked altogether from release due to the U.S. Government's process for public release or national security considerations. Although there is an overabundance of literature on the internet provided by private companies and human factors researchers, it can be challenging to verify research that has not undergone proper peer-review.

Scenarios of aviation-related complacency resulting in checklist and communication errors have been extensively researched, yet there is no clear consensus about the psychobiological reasons for the cause of complacency [21, 127]. The evidence for complacency is overwhelming, yet researchers cannot pinpoint if complacency is a cause or more of a "catch-all" of various factors leading to an accident [127]. RAND published a study focusing on stress and dissatisfaction in the USAF UAV community but only barely touched on boredom and completely disregarded complacency and habituation [93]. Satellite and RPA operations require increased research into boredom, complacency, habituation, and vigilance.

Human factors in crewed space missions have been exhaustively detailed in multiple studies regarding the China National Space Administration, European Space Agency, Indian Space Research Organization, Japan Aerospace Exploration Agency, NASA, and the Russian Federal Space Agency [62, 128]. The predominant research on remote operations shift work centers around the AFRL and the RAND Corporation's investigation into the USAF RPA community [93]. However, data from the AFRL must be extracted based on unit data to distinguish remote operations, which may not be easy to discern for those without a military background. Furthermore, the release of data products by the AFRL may lag other sources due to how the military releases information.

Aviation human factors studies are abundant due to the volume of aircraft traffic and the potential of subsequent passengers affected. However, studies do not provide a correlated picture benefiting remote operations. The literature is sparse for references on circadian rhythms human factors in space operations, yet circadian rhythms may play a significant role in satellite fault handling performance [9, 50]. Space and remote operations systems require further dedicated research to understand further the impact of shift work on the HITL and systems employed; without added research, multi-billion-dollar systems could be vulnerable to the same risks nuclear power plants and crewed space programs have experienced [9].

Cross-sector use of CRM has been proven to save lives through translatable educational communication interventions [77]. CRM has been cited as being valuable in reducing human error in space operations [104]. However, the remainder of the literature is sparse on the topic of satellite operations human factors CRM. RPA operations have benefited from advances in crewed aviation CRM training concepts. The concept of CRM extending outside the cockpit in aviation to air traffic control and other aircraft in the loop should be promoted in the satellite operations sector. CRM in satellite operations should be further researched and developed to extend outside the SOC crews on shift due to the complexities of a contested, degraded, and operationally limited orbital environment. Automation and autonomy have been referenced in multiple aviation and healthcare studies regarding bias and social loafing. Satellite systems are highly advanced, but they are not perfect; the HITL must stay vigilant and ready to respond to an anomalous event should automation or autonomy fail. Further research is required to understand the extent of automation bias, complacency, and social loafing in satellite operations centers.

Satellites on-orbit produce large quantities of data that must be downlinked to ground stations for analysis [7, 9]. Satellites tend to operate for long periods between anomalies or non-nominal events, resulting in trending data being stored for very long periods until the data is needed to be analyzed [7, 125]. Satellite telemetry data can be considered Big Data due to the size of the data, the speed at which the data transmits from a satellite to the ground station, and its relative complexity if not for the proper software infrastructure to allow for proper data analysis [129].Satellite Big Data processing may be a viable option to help mitigate human error by better equipping the HITL with the correct information ahead of time if descriptive, predictive, and prescriptive analytics are employed[124].

## 8. CONCLUSION

Human factors lessons learned are present in multiple unrelated sectors, yet actionable lessons learned have not been fully transferred to mitigate human error in remote operations. The healthcare profession has increased lifesaving capabilities using CRM principles gained from aviation, yet similarities between aviation and UAS systems have not been fully leveraged. Additionally, satellite operations have much to gain from the efficiencies and initiatives of other sectors. Keeping multi-million-dollar satellites on-orbit free from human error will take more than just acknowledging lessons learned; lessons learned must be implemented. Developers of complex remotely operated systems must consider the tendency of the HITLto seek comfort in complacency. Developers must strive to ensure automation and autonomy do not increase the risk of human error. Big Data analytics must be used in the satellite operations sector to increase HITL situational awareness in preparation for potential issues or events that would have otherwise been unknown. The case for developing novel Big Data analytics goes beyond profits, rather Big Data may be an analysis answer for mitigating human factors related errors through a better understanding of all the available data. Only through the transfer of knowledge between sectors can efficiencies be leveraged to benefit those in command of remotely operated assets.

## 9. FUTURE WORK

Future qualitative research will increase knowledge of satellite and related remote operations sectors complacency-induced error mitigation. Observational data-gathering will be undertaken to identify differences in complacency occurrences during regular and rotating shift work in space operations beyond the current state of research. Subsequent research will seek to identify a correlation between the level of system autonomy to the amount and frequency of resultant complacency-induced errors. HCI error mitigation methods will be developed and studied to mitigate complacency-induced errors. Finally, future research will identify opportunities for knowledge transfer between aerospace industry sectors to mitigate complacency-induced errors through translatable standardized prevention and mitigation methods.

### ACKNOWLEDGMENTS

David Heinrich would like to dedicate his first journal to his wife, Samantha, and kids for their unwavering support in his academic and professional pursuits. Additionally, Mr. Heinrich would like to thank his parents, Catharina and Terry Johnson, for imparting an education-first mentality and for providing their unconditional support. Mr. Heinrich would also like to thank Dr. Ian McAndrew and Dr. Jeremy Pretty for guiding him on his academic journey. Finally, Mr. Heinrich would like to pay tribute to his father, who is no longer with us: You are missed.

**AUTHORS**

**David Heinrich** is a Human Factors Ph.D. candidate at Capitol Technology University. His professional background spans over 19 years in the United States Air Force as a fighter avionics technician, instructor, and satellite operations professional. He has extensive expertise in the development and operations of war fighter-centric technologies.

**Dr. Ian McAndrew** is the Dean of Doctoral Programs at Capitol Technology University. He has taught in universities worldwide and is a frequent keynote speaker at many International Conferences. He is a Fellow of the Royal Aeronautical Society and Chartered Mechanical and Electrical Engineer in the U.K.

**Dr. Jeremy Pretty** is a Senior Program Manager with the United States Air Force Civil Service. His professional background includes over 15 years in program/project/product management within the U.S. Department of Defence on Aircraft and Information Technology systems. He is a member of the Royal Aeronautical Society.